\documentstyle[12pt]{article}

\def\TT{\ST_{[\bf{ric}]}{}  }
\def\td{\tilde}
\def\a{\alpha}
\def\b{\beta}
\def\dd{{\hbox{d}}}
\def\DD{{\hbox{D}}}
\def\g{{\gamma}} %<<<<<<<<<
\def\gma{{\gamma}}
\def\r{\rho}
\def\th{\theta}
\def\ld{........}
\def\nml{{\cal N}}
\def\ax{{\cal A}}
\def\R#1#2{R^#1{}_#2}
\def\q#1#2{q^#1{}_#2}
\def\Q#1#2{Q^#1{}_#2}
\def\Om#1#2{\Omega^#1{}_#2}
\def\om#1#2{\omega^#1{}_#2}
\def\L#1#2{\Lambda^#1{}_#2}
\def\K#1#2{K^#1{}_#2}
\def\et#1#2{\eta^#1{}_#2}
\def\e#1{e^#1{}}
\def\sp#1{#1^{\prime}}
\def\l{\lambda}
\def\gg{{\cal G}}

\def\Tor{{\bf T}}
\def\dotTor{{\bf {\dot T}}}

\def\SS{{\bf S}}
\def\dotSS{{\bf \dot S}}

\def\nab#1{\nabla_{{}_#1}}
\def\i#1{i_{{}_{#1}}}
\def\ST{{\cal T}}
\def\bfg{{\bf g}}
\def\pprime{^\prime}

\begin{document}

\title{
%%DRAFT OUTLINE May 1996\\[2cm]
NON-RIEMANNIAN\\
GRAVITATIONAL INTERACTIONS
}

\author{Robin W Tucker\\
Charles Wang\\[0.8cm]
\it School of Physics and Chemistry, Lancaster University \\
%\it University of Lancaster,\\
\it   LA1 4YB, UK\\[0.5cm]
\tt r.tucker{\rm @}lancaster.ac.uk\\
\tt c.wang{\rm @}lancaster.ac.uk
}

\maketitle
%............................... Abstract ...............................
\begin{abstract}

Recent developments in theories of non-Riemannian gravitational
interactions are outlined. The question of the motion of a fluid in the
presence of torsion and metric gradient fields is approached in terms of
the divergence of the Einstein tensor associated with a general
connection. In the absence of matter  the variational
equations associated with a broad class of actions involving non-Riemannian
fields give rise to an Einstein-Proca system associated with the
standard Levi-Civita connection.
\end{abstract}
%............................. end of Abstract ...............................

%%%%%%%%%%%%%%%%%%%%%%%%%%%%%%%%%%%%%%%%%%

\section{Introduction}

Einstein's theory of gravity has an elegant formulation in terms
of\break (pseudo-) Riemannian geometry.  The field equations follow as
a local extremum of an action integral under metric variations.  In
the absence of matter the integrand of this action is simply the
curvature scalar associated with the curvature of the Levi-Civita
connection times the (pseudo-) Riemannian volume form of spacetime.
Such a connection $\nabla$ is torsion-free and metric compatible.
Thus for all vector fields $X,Y$ on the spacetime manifold, the
tensors given by:
\begin{equation}
 \Tor(X,Y)=\nab{X}Y-\nab{Y}X-[X,Y] 
\end{equation}
\begin{equation}
\SS=\nabla \bfg 
\end{equation}
are zero,
where
 $ \bfg$ denotes the metric tensor, $\Tor$ the 2-1 torsion tensor 
and ${\SS}$ the gradient tensor of $\bfg$ with respect to $\nabla$.
Such a Levi-Civita connection provides a useful reference
connection since it depends entirely on the metric structure of the
manifold. 

Non-Riemannian geometries feature in a number of theoretical descriptions
of the interactions between fields and gravitation. Since the early pioneering
 work by Weyl, Cartan, Schroedinger and others such geometries have often
provided a succinct and elegant guide towards the search for unification of
the forces of nature
\cite{weylref} 
. In recent times interactions with supergravity have
been encoded into torsion fields induced by spinors and dilatonic
interactions from low energy effective string theories 
have been encoded into connections
that are not metric-compatible 
\cite{r1,r2,r3,r4}.
However theories in which the
non-Riemannian geometrical fields are dynamical in the absence of matter
are more elusive to interpret. It has been suggested 
that they may play an important role in certain astrophysical contexts  
\cite{newgravity}.
Part of the difficulty in interpreting such fields is that there is little
experimental guidance available for the construction of a viable
theory that can compete effectively with general relativity in domains that
are currently accessible to observation. In such circumstances one must be
guided by the classical solutions admitted by theoretical models that admit
dynamical non-Riemannian structures
\cite{Hehl,Hehl2,Baekler,Ponomariev,McCrea,RWTCW,Tey}.
This approach is being currently pursued by a number of groups
\cite{newgravity,TresA,TresB,TresC,TresD,RWTCW}. An associated problem
concerns the role of the torsion and metric gradients in determining the
motion of matter \cite{ob,neeman,autopar}. This paper briefly reports on
recent efforts to address these questions.

%%%%%%%%%%%%%%%%%%%%%%%%%%%%%%%%%%%%%%%%%%%%%%%%%%%%%%%%%%%%%%%%%%%%%%%%%%%%%%%

\def\bfR#1#2{ {  R}_{{}_{#1,#2}}  } 

\def\dotbfR#1#2{ { \dot{R}}_{{}_{#1,#2}}  } 

\def\pprime{^\prime}
\def\frac#1#2{{#1\over #2}}
\def\ot{\otimes}
\def\CC{{\cal C}}
\def\RR{{\cal R}}

\section{Non-Riemannian Geometry}

A linear connection $\nabla$ 
on a manifold provides a covariant way to differentiate
tensor fields. It provides a type preserving derivation on the algebra 
of tensor fields that commutes with
contractions.  
Given an arbitrary local basis
of vector fields $\{X_a\}$ the most general
 linear  connection is specified locally
by a set of $n^2$ 1-forms  $\Lambda^a{}_b$ where $n$ is the dimension of the
manifold:
\begin{equation}
\nab{{X_a}} \,X_b=\Lambda^c{}_b (X_a)\, X_c . 
\end{equation}
Such a  connection can be fixed by specifying a (2, 0)
 symmetric metric tensor
$\bfg$,  a (2-antisymmetric, 1) tensor $\Tor$ and a (3, 0) tensor 
$\SS$, symmetric
in its last two arguments. If we require that $\Tor$ be the torsion of
$\nabla$ and $\SS$ be the gradient of $\bfg$ then it is straightforward to
determine the connection in terms of these tensors. Indeed since $\nabla$
is defined to commute with contractions and reduce to differentiation on
scalars it follows from the relation
\begin{equation}
X(\bfg(Y,Z))=\SS(X,Y,Z)+\bfg(\nab{X}Y,Z)+\bfg(Y,\nab{X}Z)
\end{equation} 
that 
\begin{eqnarray}
2\bfg(Z,\nab{X}Y)&=&X(\bfg(Y,Z))+Y(\bfg(Z,X))-Z(\bfg(X,Y))
\nonumber\\
&&{}-\bfg(X,[Y,Z])-\bfg(Y,[X,Z])-\bfg(Z,[Y,X])
\nonumber\\
&&{}-\bfg(X,\Tor(Y,Z))-\bfg(Y,\Tor(X,Z))-\bfg(Z,\Tor(Y,X))
\nonumber\\
&&{}-\SS(X,Y,Z)-\SS(Y,Z,X)+\SS(Z,X,Y)
\end{eqnarray}
where $X,Y,Z$ are any  vector fields.
The general curvature operator 
%%%%$\bfR{X}{Y}$ 
${\bf R}_{X,Y}$
defined in terms of $\nabla$
by
\begin{equation}
{\bf R}_{X,Y}Z=\nab{X}\nab{Y}Z-\nab{Y}\nab{X}Z-\nab{{[X,Y]}}Z 
\end{equation}
is also a type-preserving tensor derivation on the algebra of tensor fields.
The general (3, 1) curvature tensor ${\bf R}$ of $\nabla$ is defined by
\begin{equation}
{\bf R}(X,Y,Z,\beta)=\beta(\bfR{X}{Y} Z)
\end{equation}
where $\beta$ is an arbitrary 1-form. This tensor gives rise to a set of
local curvature 2-forms $R^a{}_b$: 
\begin{equation}
R^a{}_b(X,Y)=\frac{1}{2}\,{\bf R}(X,Y,X_b,e^a)
\end{equation}
where $\{e^c\}$ is any local basis of 1-forms dual to $\{X_c\}$.
In terms of the contraction operator $i_X$ with respect to $X$
one has $i_{X_b}\,e^a\equiv
i_b\,e^a=e^a(X_b)=\delta^a{}_b$.
In terms of the connection forms
\begin{equation}
\R{a}{b}=\dd \Lambda^a{}_b+\Lambda^a{}_c\wedge \Lambda^c{}_b .
\end{equation}
In a similar manner the torsion tensor gives rise to a set of local torsion
2-forms $T^a$:
\begin{equation}
T^a(X,Y)\equiv \frac{1}{2}\, e^a(\Tor(X,Y)) 
\end{equation}
which can be expressed in terms of the connection forms as
\begin{equation}
T^a=\dd e^a+\Lambda^a{}_b\wedge e^b.
\end{equation}
Since the metric is symmetric the tensor $\SS$ can be used to define a set
of local non-metricity 1-forms $Q_{ab}$ symmetric in their indices: 
\begin{equation}
Q_{ab}(Z)=\SS(Z,X_a,X_b). 
\end{equation}

%%%%%%%%%%%%%%%%%%%%%%%%%%%%%%%%%%%%%%%%%%%%%%%%%%%%%%%%%%%%%%%%%%%%%%%%%%%%%

\section {The Motion of a Charged Fluid}

\def\und{\underbrace}

\def\TT{\ST_{[{\bf ric}]}{}  }

\def\td{\tilde}
\def\a{\alpha\,}
\def\b{\beta\,}
\def\dd{{\hbox{d}}}
\def\DD{{\hbox{D}}}
\def\gma{{\gamma}}
\def\g{{\mbox{\boldmath $\gamma$}}}
\def\TT{{\mbox{\boldmath $\tau$}}}
\def\Cal{\cal}
\def\r{\rho}
\def\th{\theta}
\def\ld{........}
\def\nml{{\cal N}}
\def\ax{{\cal A}}
\def\R#1#2{R^#1{}_#2}
\def\q#1#2{q^#1{}_#2}
\def\Q#1#2{Q^#1{}_#2}
\def\Om#1#2{\Omega^#1{}_#2}
\def\om#1#2{\omega^#1{}_#2}
\def\L#1#2{\Lambda^#1{}_#2}
\def\K#1#2{K^#1{}_#2}
\def\et#1#2{\eta^#1{}_#2}
\def\e#1{e^#1{}}
\def\sp#1{#1^{\prime}}
\def\l{\lambda}
\def\gg{{\cal G}}
\def\Tor{{\bf T}}
\def\dotTor{{\bf {\dot T}}}

\def\SS{{\bf S}}
\def\dotSS{{\bf \dot S}}
\def\nab#1{\nabla_{{}_#1}}
\def\i#1{i_{{}_{#1}}}
\def\ST{{\cal T}}

\def\bfR#1#2{ {\bf R}_{{}_{#1,#2}}  } 
\def\dotbfR#1#2{ {\bf \dot{R}}_{{}_{#1,#2}}  } 
\def\bfg{{\bf g}}
\def\pprime{^\prime}
\def\frac#1#2{{#1\over #2}\,}
\def\ot{\otimes}
\def\CC{{\cal C}}
\def\k{\kappa\,}

\def\dott{\dot{\,}}
\def\dQ{\dd\,Q}
\def\var#1#2{ \und{(#1) }_{#2}\dott}

\def\Ein{{\bf Ein}}
\def\Ric{{\bf Ric}}
\def\Richat{{\bf \widehat{Ric}}}
\def\bfg{{\bf g}}

The traditional
 Einstein-Maxwell equations for an electrically  charged ideal fluid
with mass density $\rho$ 
(and composed of elements with charge to mass ratio $\frac{e}{m}$)
are~:
\begin{equation}
\Ein={\TT}\label{EinMax}
\end{equation}
\begin{equation}
\dd * F=j
\end{equation}
\begin{equation}
\dd F=0
\end{equation}
 where
$\Ein=\Ric-\frac12\,\bfg\RR$ is the Einstein tensor associated with the
Levi-Civita connection $\nabla$ and 
${ \TT}=\rho\widetilde V \otimes \widetilde V + {\TT}_{Max}$ is given 
in terms of the stress-energy tensor $ {\TT}_{Max}$ of the Maxwell 2-form
$F$. 
The Hodge map associated with $\bfg$ is denoted by $*$.
The electric current 3-form
$j=*(\rho_e\widetilde V)$ is specified 
in terms of the  charge density $\rho_e=\frac{e}{m}\rho$ 
of the fluid with  1-form velocity $\widetilde V$.
Since $\nabla . \Ein=0$ and $\dd\,j=0$ it follows from Eq.~(\ref{EinMax})  that
$\nabla . \TT=0$ or
\begin{equation}
\nabla_V\, \widetilde V=-\frac{e}{m}{{i_V} F}.
\label{Lorentz}
\end{equation}
Thus each integral curve of the vector field 
$V$ models the world line of an electrically charged
fluid element under the influence of the Lorentz force. The gravitational
forces are encoded into the Levi-Civita connection and are determined by the
metric tensor alone.

In a non-Riemannian theory of gravitation the additional torsion and metric
gradient fields are expected to provide additional forces, the nature of
which depend on the field equations.  

%%%%%%%%%%%%%%%%%%%%%%%%%%%%%%%%%%%%%%%%%%%%%%%%%%%%%%%%%%%%%%%%%%%%%%%%%%%%%%

\section{Divergence of the Generalised Einstein Tensor}

\def\M{{\bf M}}
\def\R{{\bf R}}
\def\r{{\cal R}}

\def\B{{\cal B}}
\def\A{{\cal A}}
\def\sig{{\cal V}}
\def\xx{{\cal W}}

\def\S{{\bf S}}

\def\div{\nabla{}.{}}

Suppose the non-Riemannian Einstein field equations are given by
Eq.~(\ref{EinMax}) for some general stress-energy tensor $\TT$ 
where
\begin{equation}
\Ein\equiv\Richat -\frac12 \bfg \r
\label{NRE}
\end{equation}
is given in terms of the Ricci tensor
$\Ric (X,Y)=\R(X_a,X,Y,e^a)$ by
\begin{equation}
\Richat(X,Y)=\frac12(\Ric(X,Y)+\Ric(Y,X)).
\end{equation} 
Unlike the Einstein tensor associated with the Levi-Civita connection,
the Einstein tensor $\Ein$  defined in Eq.~(\ref{NRE}) is 
associated with the  non-Riemannian 
connection and is not in general  divergenceless:
\begin{equation}
(\div \Ein)(X_b) \equiv (\nabla_{X_a}\Ein)(X^a,X_b)
\neq 0.
\end{equation}
The departure from zero should be expressible in terms of the non-Riemannian
fields $\Tor$, $\SS$, contractions of the generalised curvature
tensor and their covariant derivatives. Like the
electrical Lorentz force in Eq.~(\ref{Lorentz})  above such terms may be 
associated with 
{\it gravitational}
 forces that in general may produce non-geodesic motion of matter.
Thus an indication of the motion of  a fluid in a general non-Riemannian 
gravitational is given by the divergence of Eq.~(\ref{EinMax}):
\begin{equation}
\nabla . \TT=\nabla . \Ein
\label{EqnMotion}
\end{equation}
where $\TT$ contains the stress-energy  tensor of the fluid.

The non-Riemannian forces $\nabla . \Ein$ may be expressed
as follows. Introduce the tensors
\begin{equation}
{\bf ric} (X,Y)=\R(X,Y,X_a,e^a)
\end{equation}
\begin{equation}
{\cal M}\equiv \e c\{\und{\cal S}_{X,Y,X_c}
\big ((\nabla_{X_c})(X,Y)-T(X_c,T(X,Y))  \big )\}
\end{equation}
%%$$ M=\frac12 ({\cal M}-{\bf ric} )$$
where $\und{\cal S}_{X,Y,Z}$ denotes a cyclic sum over $X,Y,Z$.
It follows that
%$$\Richat=\Ric-\frac12 ({\cal M}-{\bf ric} ).$$
\begin{equation}
\Richat=\Ric-\M
\end{equation}
where $\M\equiv \frac12 ({\cal M}-{\bf ric} )$.
For any tensor
${\cal T}$ and vector fields $X,Y,Z,U,$\break $V,W,\ldots $, define the tensor 
${\cal T}(X,Y)$
 by
\begin{equation}
({\cal T}(X,Y))(U,W,\ldots)\equiv{\cal T}(X,Y,U,W,\ldots).
\end{equation}
Furthermore for any vector $X$  denote its metric dual by $\widetilde X$
where $\widetilde X=\bfg(X,-)$.
Similarly for any 1-form $\beta$ denote its metric dual by $\widetilde \beta$.
In terms of this notation one first computes
\begin{equation}
(\nabla_{X_a}\Richat)(X^a,X_b)\equiv(\div \Richat)(X_b)
=(\div\Ric -\div\M)(X_b).
\end{equation} 
 From the definitions above one finds
\begin{equation}
(\div \Ric)(X_b)=\frac12 X_b\r +\B(X_b)
\end{equation}
where
\begin{eqnarray}
\B(X)&\equiv&\frac12 \S(X,X^b,X^c)\Ric(X_c,X_b)-2(\nabla_{X_a}\M)(X,X^a)
\nonumber\\
&&{}+\frac12\sig(X_a,X,X_\gamma,X^b,e^a)-
\frac12\sig(X,X_a,X_b,X^a,e^b)
\end{eqnarray}
and
\begin{eqnarray}
\sig(W,X,Y,Z,\b)&\equiv&
\R(X,Y,\widetilde\b,\S(W,Z,-))-
\R(X,Y,\widetilde \S(W,\b,-),\widetilde Z)
\nonumber\\
&&{}-\xx(X,Y,Z,\widetilde \S(W,\b,-))+
(\nabla_W\xx)(X,Y,Z,\widetilde \b)\nonumber\\
\end{eqnarray}
where
\begin{equation}
\xx(U,W)\equiv(\nabla_U \S)(W)-(\nabla_W \S)(U)+\S(T(U,W)).
\end{equation}
In terms of $\div\M\equiv(\nabla_{X_a}\M)(X^a,-)$,
it follows that
\begin{equation}
\div\Richat =\frac12 \dd\,\r +\B-\div\M
\label{dv}
\end{equation}
where
\begin{equation}
\r\equiv \Ric(X_a,X^a).
\end{equation}
Next one calculates
$\div(\bfg\r)=(\nabla_{X_a}(\bfg \r))(X^a,-)$. Since
\begin{equation}
\nabla_{X_a}(\bfg \r)=\r\,\S(X_a,-,-)+X_a\r\,\,\bfg
\end{equation}
then
\begin{equation}
\div(\bfg\,\r)=\r\,\S(X_a,X^a,-)+\dd\,\r .
\label{dvv}
\end{equation}
Putting together Eq.~(\ref{dv}) and Eq.~(\ref{dvv}) one obtains
\begin{equation}
\div\Ein=\B-\div\M-\frac12 \r\,\S(X_a,X^a,-).
\label{divEin}
\end{equation}
Inserting Eq.~(\ref{divEin}) in Eq.~(\ref{EqnMotion}) yields an equation of
motion for the fluid
leading to an interpretation of
$\div\Ein$ in terms of forces on the matter described by $\TT$.

\section{Non-Riemannian Actions}

\newcommand{\nno}{\mbox{\nonumber}}

%******************************************************************************
\def\nablc#1{\widehat{\nabla}_{{}_#1}}
\def\lamb{{\mbox{\boldmath $\lambda$}}}
\def\EinLC{{\bf Ein}\kern-1em\raise0.4em\hbox{$^\circ$}\kern0.6em}
\def\RicLC{{\bf Ric}\kern-1em\raise0.4em\hbox{$^\circ$}\kern0.6em}
\def\delLC{\hbox{$\nabla$}\kern-0.6em\raise0.4em\hbox{$^\circ$}\kern0em}
\def\DDLC{\hbox{D}\kern-0.6em\raise0.4em\hbox{$^\circ$}\kern0em}
\def\RLC{\hbox{$\cal R$}\kern-0.6em\raise0.4em\hbox{$^\circ$}\kern0em}
\def\LL{{\cal L}}
\def\QQ{{\cal Q}}
\def\FF{{\cal F}}
\def\E{{\cal E}}
\def\TTT{{\cal T}}
\def\bfh{{\bf h}}
\def\hlambda{{\widehat{\lambda}}}
\newcommand{\itLamb}{\mbox{${\mit \Lambda}$}}

To gain further insight into the nature these non-Riemannian forces one
must exploit the remaining field equations that determine the torsion and
metric gradients.
These  are most naturally
derived from an action principle in which the metric, components of
the connection and matter fields
are the configuration variables. One requires that an action
be stationary with respect to suitable variations of such variables. 
If the action $n$-form
%$\LL(\bfg,\nabla,\cdots)$ in 
$\itLamb(\bfg,\nabla,\cdots)$ in 
$n$ dimensions  contains the Einstein-Hilbert form
\begin{equation}
%\LL_{EH}(\bfg,\nabla)=\RR *1
\itLamb_{EH}(\bfg,\nabla)=\RR *1
\label{EH}\end{equation}
%and $\underbrace{\dot{\LL_{EH}}}_{\bfg}$ denotes the variational derivative
%of $\LL_{EH}(\bfg,\nabla)$ with respect to $\bfg $
and $\underbrace{\dot{\itLamb_{EH}}}_{\bfg}$ denotes the variational derivative
of $\itLamb_{EH}(\bfg,\nabla)$ with respect to $\bfg $
then
\begin{equation}
%\underbrace{\dot{\LL_{EH}}}_{\bfg}=
\underbrace{\dot{\itLamb_{EH}}}_{\bfg}=
-h^{ab}\,\Ein(X_a,X_b)\,*1
\end{equation}
where $h_{ab}\equiv\dot{\bfg}(X_a,X_b)$.
It is instructive to decompose the non-Riemannian Einstein tensor
into parts that depend on the Levi-Civita connection $\delLC$. For this
purpose  introduce the tensor $\lamb$  by
\begin{equation}
\lamb(X,Y,\beta)=\beta(\nabla_{X}{Y})-\beta(\delLC_{X}{Y})
\label{lambdef}\end{equation}
for arbitrary vector fields $X,Y$ and 1-form $\beta$.
In terms of the exterior covariant derivative ${\DDLC}$
and  Ricci tensor $\RicLC$ associated with the  Levi-Civita connection 
one may write:
\begin{equation}
\Ric(X_a,X_b)=\RicLC(X_a,X_b)
+i_{a}\,i_{c}\,(
\DDLC\, \lambda{}^c{}_{b}+\lambda{}^c{}_{d} \wedge\lambda{}^d{}_{b})
\end{equation}
where $
\lambda{}^a{}_{b}\equiv\lamb(-,X_b,e^a)
$ is a set of local 1-forms.   
It follows from Eqn.\ref{NRE} that $\Ein$ differs from
the  Levi-Civita Einstein tensor $\EinLC$  by terms involving the
tensor $\lamb$ and its derivatives.
However for a large class of actions containing the torsion and metric
gradient fields one finds that these terms can be dramatically simplified.

To see this simplification most easily it is preferably to change
 variables from $\bfg , \nabla$ to $\bfg , \lamb$ in the total action
 $n$-form and write $
%\LL(\bfg,\nabla,\cdots)=$\break $\LL'(\bfg,\lamb,\cdots)
\itLamb(\bfg,\nabla,\cdots)=$\break $\LL(\bfg,\lamb,\cdots)
$. Since 
%$\underbrace{\dot{{\delLC}}}_\nabla=0$ 
${\delLC}$ depends only on the metric
it follows from Eqn.\ref{lambdef}
that 
$
\underbrace{\dot{\lamb}}_\nabla=\dot{\nabla}
$
and
$
\underbrace{\dot{\lamb}}_\bfg=
-\underbrace{\dot{\delLC}}_\bfg
$.
Hence the variational field equations are 
\vskip -10pt
\begin{equation}
%\underbrace{\dot{\LL}}_{\nabla}=
%\underbrace{\dot{\LL'}}_\lamb=0
\underbrace{\dot{\itLamb}}_{\nabla}=
\underbrace{\dot{\LL}}_\lamb=0
\label{delvar}\end{equation}

\begin{equation}
%\underbrace{\dot{\LL}}_{\bfg}=
%\underbrace{\dot{\LL'}}_\bfg
%+\underbrace{\dot{\lamb}}_\bfg \underbrace{\dot{\LL'}}_\lamb=0
\underbrace{\dot{\itLamb}}_{\bfg}=
\underbrace{\dot{\LL}}_\bfg
+\underbrace{\dot{\lamb}}_\bfg \underbrace{\dot{\LL}}_\lamb=0
\label{gvar}\end{equation}
>From Eqn.\ref{delvar} one sees that
 the term $\underbrace{\dot{\LL}}_\lamb$ 
in Eqn.\ref{gvar} does not contribute.
To evaluate these variations one first expresses the action in terms of $\bfg$
and the components of $\lamb$ and its derivatives. To this end it is
convenient to introduce the (traceless) 1-forms 
$
\hlambda^a{}_{b}\equiv\lambda{}^a{}_{b}
-{1\over n}\,\delta{}^a{}_{b}\,\lambda{}^d{}_{d}
$, and the (traceless) 0-forms
$
\hlambda^a{}_{bc}\equiv\lambda{}^a{}_{bc}
-{1\over n}\,\delta{}^a{}_{b}\,\lambda{}^d{}_{dc}
$
where $\lambda{}^a{}_{bc}\equiv i_c\,\lambda{}^a{}_{b}$.
If we express the total action $n$-form $\LL(\bfg,\lamb, \cdots)$
as
\begin{equation}
\LL(\bfg,\lamb, \cdots)=\LL_{EH}(\bfg,\lamb)+\FF(\bfg,\lamb, \cdots)
\label{action}\end{equation}
for some form $\FF(\bfg,\lamb, \cdots)$
where
\begin{equation}
\LL_{EH}=\RLC-\hlambda^a{}_c \wedge \hlambda^c{}_b \wedge * (e^b \wedge e_a)
-\dd(\hlambda^a{}_b \wedge *(e^b \wedge e_a))
\end{equation}
in terms of the Levi-Civita scalar curvature $\RLC$\ \ then
\begin{equation}
\underbrace{\dot{\LL_{EH}}}_\bfg
=-h^{ab}\,\EinLC(X_a,X_b)\,*1-h^{ab}\,\E_{ab}
\quad\quad\hbox{(mod d)}
\end{equation}
where
\begin{equation}
\E_{ab}=
{1\over2}\,\hlambda^q{}_d \wedge \hlambda^d{}_p \wedge
\{
g_{ab} * (e^p \wedge e_q) 
- \delta^p{}_a * (e_b \wedge e_q)
- \delta^p{}_b * (e_a \wedge e_q)
\}.
\label{stuff}\end{equation}
Thus the Einstein field equation is
\begin{equation}
\EinLC(X_a,X_b)\,*1+\E_{ab}+\TTT_{ab}=0
\end{equation}
where $
\underbrace{\dot{\FF}}_{\bfg}=
-{h}^{ab}\,\TTT_{ab}$.
Next the variations with respect to $\lamb$ yield
\vskip -10pt
\begin{equation}
\underbrace{\dot{\LL_{EH}}}_\lamb
=\dot{\hlambda}{}^a{}_b \wedge
\{\hlambda^c{}_a \wedge * (e^b \wedge e_c)
-\hlambda^b{}_c \wedge * (e^c \wedge e_a)\}
\quad\quad\hbox{(mod d)}
\end{equation}
\vskip -10pt
\begin{equation}
\underbrace{\dot{\FF}}_{\lamb}=
\dot{\lambda}^a{}_{b}\wedge \FF^{b}{}_a.
\end{equation}
By splitting off the trace part of the resulting field equation
%$\{\hlambda^c{}_a \wedge * (e^b \wedge e_c)
%-\hlambda^b{}_c \wedge * (e^c \wedge e_a)\}+\FF^b{}_a=0$ 
one may write:
\begin{equation}
\FF^a{}_a=0
\label{cart1}\end{equation}
\begin{equation}
\hlambda^c{}_a \wedge * (e^b \wedge e_c)
-\hlambda^b{}_c \wedge * (e^c \wedge e_a)+\widehat{\FF}^b{}_a=0
\label{cart2}\end{equation}
where $
\widehat{\FF}^a{}_b\equiv {\FF}^a{}_b - {1\over n}\delta^a{}_b\,\FF^c{}_c$.
To illustrate how the field equations 
Eqn.\ref{cart1} and Eqn.\ref{cart2} can greatly simplify
the terms in Eqn.\ref{stuff}
 consider the eight parameter class of models in which
the torsion and metric gradient fields enter the action according to:
\begin{eqnarray}
\FF&=&{4\,\sigma}\,R^a{}_a \wedge *R^b{}_b
-2\,\ell\,*1+\alpha_1\,Q \wedge *Q
\nno\\ 
&&
\hspace{-20pt}
+\alpha_2\,u \wedge *u
+\alpha_3\,v \wedge *v
+\alpha_4\,Q \wedge *u
+\alpha_5\,Q \wedge *v
+\alpha_6\,u \wedge *v
\label{FF}
\end{eqnarray}
where $\sigma,\alpha_k$ are arbitrary coupling constants and $\ell$ is a
cosmological constant.
The 1-forms $u$ and $v$ may be expressed in terms of the torsion forms $T^a$
and non-metricity forms $Q_{ab}$ as follows:
\begin{equation}
u\equiv \lambda^c{}_{ac}\,e^a=T-{1\over2}\,Q
\end{equation}
\begin{equation}
v\equiv\lambda_a{}^c{}_c\,e^a={1\over2}\,Q-{1\over2}\,\QQ-T
\end{equation}
where $
T\equiv i_{a}T^a
$, $Q\equiv Q^a{}_a=-2\,\lambda^a{}_a$
and $
\QQ\equiv e^a\,i^{b} Q_{ab}
$.
Furthermore $R^a{}_a=-{1\over2}\dd Q$ is proportional to the Weyl field
2-form $\dd Q$.
Computing the variational derivatives above one finds that Eqn.\ref{cart1}
yields:
\begin{equation}
\dd * \dd Q
+{4n\alpha_1-\alpha_4-\alpha_5\over4n\sigma}\, * Q
+{2n\alpha_4-2\alpha_2-\alpha_6\over4n\sigma}\, * u
+{2n\alpha_5-2\alpha_3-\alpha_6\over4n\sigma}\, * v
=0\label{preproca}
\end{equation}
while Eqn.\ref{cart2} implies:
\begin{equation}
u=\beta_1\,Q
\label{u}\end{equation} 
\begin{equation}
v=\beta_2\,Q
\label{v}\end{equation}
and
$$\hlambda^a{}_{bc}=
-{n\beta_1+n\beta_2+1\over n(n-1)(n-2)}\,\delta^a{}_b\,i_c\,Q
$$
$$
+{2\beta_2-2\beta_1+2n\beta_2+1\over 2(n-1)(n-2)}\,g_{bc}\,i^a Q
+{2\beta_2-2\beta_1+2n\beta_2+1\over 2(n-1)(n-2)}\,\delta^a{}_c\,i_b Q
$$
where
$$
\beta_{{1}}=
$$$$
\{
n (n-2 )-4 (n+2 ) (n-1
 )^{2}\alpha_{{3}}\alpha_{{4}}+2 (n+2
 ) (n-1 )^{2}\alpha_{{5}}\alpha_{{6}}
-2n (n-1 )\alpha_{{3}}
+ (4-4n 
)\alpha_{{4}}
$$$$
+2 (n-1 ) ({n}^{2}-2
 )\alpha_{{5}}+n (n-1 )\alpha_{{6}}
\}\{
-2{n}^{2} (n-2 )
-2 (n+2 )
 (n-1 )^{2}{\alpha_{{6}}}^{2}
$$$$
+8 (n+2
 ) (n-1 )^{2}\alpha_{{2}}\alpha_{{3}}+
 8(n-1)\alpha_{{2}}+ 8(n-1)\alpha_{{3}}-4 (n-1 ) ({n}^{2
}-2 )\alpha_{{6}}
\}^{-1}
$$
$$
\beta_{{2}}=
$$$$
\{
n (n-2 )+2 (n+2 ) (n-1
 )^{2}\alpha_{{4}}\alpha_{{6}}-4 (n+2
 ) (n-1 )^{2}\alpha_{{2}}\alpha_{{5}}-
2n (n-1 )\alpha_{{2}}
$$$$
+2 (n-1 
) ({n}^{2}-2 )\alpha_{{4}}
+ (4-4n
 )\alpha_{{5}}+n (n-1 )\alpha_{{6}}
\}\{
-2{n}^{2} (n-2 )-2 (n+2 )
 (n-1 )^{2}{\alpha_{{6}}}^{2}
$$$$
+8 (n+2
 ) (n-1 )^{2}\alpha_{{2}}\alpha_{{3}}
+8(n-1)\alpha_{{2}}+ 8(n-1)\alpha_{{3}}-4 (n-1 ) ({n}^{2
}-2 )\alpha_{{6}}\}^{-1}.
$$
Substituting Eqn.\ref{u} and  Eqn.\ref{v} into
Eqn.\ref{preproca}   one now obtains a Proca-type equation in the
form
\begin{equation}
\dd * \dd Q
+\beta_3\, * Q
=0
\label{proca}
\end{equation}
%\vfill\eject
where
\begin{equation}
\beta_{{3}}=
{4n\alpha_1-\alpha_4-\alpha_5\over4n\sigma}\, 
+{2n\alpha_4-2\alpha_2-\alpha_6\over4n\sigma}\,\beta_{{1}} 
+{2n\alpha_5-2\alpha_3-\alpha_6\over4n\sigma}\, \beta_{{2}}
\end{equation}
From the metric variation of $\FF$ one finds
$$
\TTT_{ab}=
\ell\,g_{ab}*1+
\sigma\,\TT(X_a,X_b)*1
+\widehat{\TTT}_{ab}
$$
where
$$
\TT=*^{-1}
\{
\beta_3\,(i_a Q \wedge * i_b Q -{1\over2}\,g_{ab}\, Q \wedge * Q)
+i_a \dd Q \wedge * i_b \dd Q -{1\over2}\,g_{ab}\, \dd Q \wedge * \dd Q\}\,
e^a \otimes e^b
$$
and
$$
\widehat{\TTT}_{ab}=
\zeta_1\,
g_{ab} Q \wedge * Q 
+
\zeta_2\,i_a Q \wedge * i_b Q
$$
in terms of
$\zeta_1={1\over8}\{ (n+1 ) (n-2 )+4 (n-2
 ) ({n}^{3}+{n}^{2}-4n-6 )\beta_{{1}
}\beta_{{2}}$\break
$-(2{n}^{2}+8-2{n}^{3}+8n
 )\beta_{{2}}
- (24+12n-4{n}^{2})
{\beta_{{2}}}^{2}
+ (2{n}^{3}-2{n}^{2}+8 
)\beta_{{1}}$\break
$+ (12{n}^{2}-24+4n ){\beta_{
{1}}}^{2}\}\,(n-1)^{-2}(n+2)^{-2}\,$
and
$\zeta_2=-{1\over4}\{
n-2-8 (n+1 )$\break
$(n-2 )\beta_{{1}}
\beta_{{2}}-8\beta_{{2}}n-(8+4{n}^{2}+4n
 ){\beta_{{2}}}^{2}+4\beta_{{1}}{n}^{2}
-(
8+4n-4{n}^{2}-4{n}^{3} ){\beta_{{1}}}^{2}\}$\break
$\,(n-1)^{-2}(n+2)^{-2}.$
Remarkably $\widehat{\TTT}_{ab}$ exactly cancels $\E_{ab}$ leaving the
Einstein equation in the form
\begin{equation}
\EinLC+\ell\,\bfg+\sigma\,\TT=0
\label{ein}\end{equation}
For $n=4$ the above framework has been used \cite{tuc_wang} 
 to confront the problem of
dark matter gravitational interactions in the context of inflationary
cosmolgy and Newtonian galactic dynamics.

\section{Conclusion}
The solutions to the field equations derived from
the action
$\LL_{EH}(\bfg,\lamb)+\FF$
where $\FF$ is given by
 Eqn.\ref{FF}  may be generated from solutions to the 
 Levi-Civita Einstein-Proca
system Eqn.\ref{proca}, Eqn.\ref{ein} with arbitrary cosmological constant 
$\ell$. 
Such a reduction promises considerable simplification
in the description of the behaviour of matter in the presence
 of non-Riemannian gravitational fields.

%*****************************************************************************

\section*{Acknowledgement}

The authors are grateful to J Schray for valuable interaction. 
 RWT is grateful   
to the Human Capital and Mobility Programme of the European Union for
partial support. CW is grateful to the Committee of Vice-Chancellors and 
Principals, UK for an Overseas Research Studentship and to
Lancaster University for a School of Physics and Chemistry
Studentship and a Peel Award.

\end{document}